# Neutron Dose measurement in the carbon ion therapy area at HIRFL (IMP) as $^{12}$C ions with energies from 165 to 350MeV/u

XU Jun-Kui[1,2], LI Wu-Yuan[1], Yan Wei-Wei [1]，SU You-Wu[1†], Li Zong-Qiang[1], Mao Wang[1], PANG Cheng-Guo [1,2], XU Chong[1]

1 Institute of Modern Physics, Chinese Academy of Science, Lanzhou 730000, China
2 School of Nuclear Science and Technology, Lanzhou University, Lanzhou 73000, China

**Abstract**：The neutron dose distributions on observation distances and on observation angles were measured using a Wendi-II neutron dose-meter at the deep tumor therapy terminal at HIRFL (Heavy Ion Research Facility in Lanzhou) as $^{12}$C ions with energies from 165 to 350 MeV/u bombarding on thick solid water targets with different thickness according to the ion energies. The experimental results were compared with those calculated by FLUKA code. It is found that the experimental data was in good agreement with the calculated results. The neutron energy spectra were also studied by using the FLUKA code. The results are valuable for the shielding design of high energy heavy ion medical machines and for the individual dose assessment.

## 1 Introduction

Carbon ions are the most common particles in heavy ion tumor therapy due to their physical and biological properties. The so called "Bragg peak" of heavy ions makes it very suitable for radiotherapy, but secondary particles caused by heavy ion reactions must be carefully considered in the shielding design of high energy heavy ion medical machines and treatment planning models. Neutrons are the most abundant products in all secondary particles and very important in safety evaluation for heavy ion therapy, although its dose in tissue is fairly small due to it cannot cause ionization directly. This is because the neutrons can affect a large area due to its strong penetrating power, that is to say, it can influence the whole body of the patient including tumor and health tissue.

In our previously work, neutron dose distributions have been measured in the surface tumor treatment terminal at HIRFL [1]. In that experiment, carbon ions were accelerated to 100 MeV/u by SSC of HIRFL, and then delivered vertically down to a basement where the surface tumor treatment terminal located. Since the maximum energy can be accelerated by SSC is 100 MeV/u for $^{12}$C, it is not enough for deep tumor therapy. Therefor HIRFL-CSR is used to accelerate the $^{12}$C ions to higher energies (maximum energy is 900 MeV/u) for studying the deep tumor therapy. Fig. 1 shows the layout of the HIRFL-CSR and the location of the deep tumor therapy terminal.

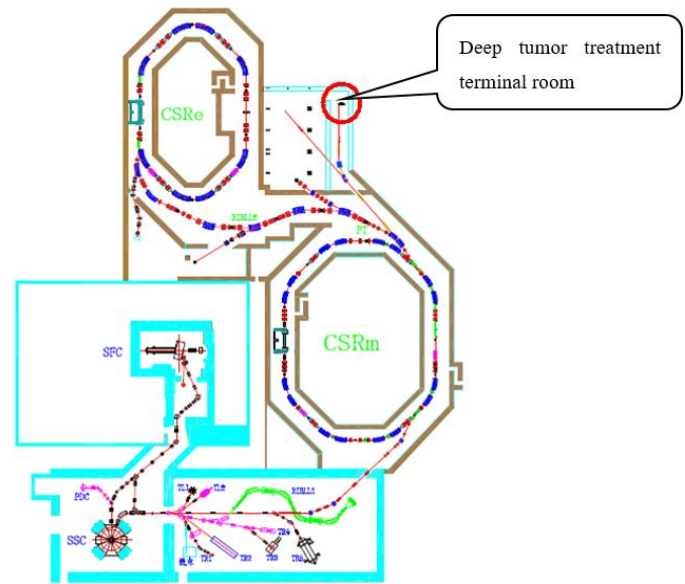

Fig. 1. Schematic diagram of the HIRFL-CSR.

Studies of the fragmentation of light ions in water or tissue-substitute materials in biomedical applications can cast back to 1971 at Princeton by W. Schimmerling *et al.* [2]. Detailed measurements of fragmentation of 670 MeV/u $^{20}$N hitting on water target were performed by W. Schimmerling *et al.* [3] at the BEVALAC (LBL). Other reports on nuclear fragmentation of heavy ions in biomedical applications can be found till very recently [4, 5, 6].

Neutron yields, energy spectra and angular distributions induced by mid-energy heavy-ions bombarding on thick targets have been measured intensively in past decade years. G. Li *et al.* measured neutron yields by 50-100 MeV/u heavy ions hitting on thick targets using activation method at HIRFL IMP [7]. T. Kurosawa *et al.* performed a systematic experimental study on neutron yields, spectra and angular distributions of various ions on thick targets with TOF method at HIMAC (Heavy Ion Medical Accelerator in Chiba) [8, 9], and compared their experiments with the HIC code calculation results. These studies are useful to radiation protection such as neutron shielding design, and have significance in tumor therapy.

Differing from electrons and gamma rays, the relative biological effectiveness (RBE) of neutrons strongly depends on their energies. An ideal neutron dose detector should have the same response as human body, but unfortunately we cannot get such detector up to now. For monitoring neutron dose a so called A-B rem-meter [10] detector is widely used, which is made of the Bonner Ball with a thermal neutron absorber inside. This kind of detectors has a good energy response to neutrons with energies up to about 20 MeV. However, for higher energy neutrons it underestimates seriously the neutron dose. Thus some modification to such standard A-B remmeter have been made. Calculations show that the modifications can increase the neutron response in high energies while keeping the response curve unchanged in the lower energy region [11, 12, 13]. However, it is still a problem in calibrating the detector at high energies due to the lack of the mono-energetic neutron sources, thus it has some uncertainty in measuring the neutron dose. If we have the neutron spectra it is possible to obtain the neutron dose by correcting the measured results with a standard A-B rem-meter.

In the present work, the neutron dose was measured as carbon ions with specific energy of 165, 207, 270 and 350 MeV/u bombarding on thick solid water target. Then we calculated neutron dose distribution and energy spectra with FLUKA code. It is found that the measured neutron dose results have a good agreement with the calculated results.

## 2 Experiments

The experiment was carried out at the deep tumor therapy terminal of HIRFL-CSR. $^{12}$C ions were accelerated to 80 MeV/u by SSC then injected into CSRm for accelerating to energies needed, finally the ions were extracted and delivered to the deep tumor therapy terminal chamber, in which a solid water target with thickness enough to stop the ions was placed on the treatment bed. A Wendi-II neutron dose-meter was placed at the same level as the target (1.4 meter from the floor, fig.2) for detecting the produced neutron dose. Table 1 lists the material composition of the target. The target used in experiments is the superposition of several thin targets, whose cross section is 30 cm × 30 cm. Table 2 shows the thickness of the solid water target used for different energy carbon ions incidence. Neutrons emitted from the target were measured by the dose-meter at different angles and distances. The beam intensity was measured by a plate ionization chamber which located behind the beam extraction window. The ionization chamber is filled with the nitrogen gas with an atmosphere pressure.

Table 1. Material composition of the solid water target, the density is 1.043 ±0.005 g/cm3.

| Component | H | C | N | O | Cl | Ca |
|---|---|---|---|---|---|---|
| Proportion (%) | 8.1 | 67.2 | 2.4 | 19.9 | 0.1 | 2.3 |

Table 2. The solid target thickness with different carbon ion energies.

| Ion Energy (MeV/u) | 165 | 207 | 270 | 350 |
|---|---|---|---|---|
| Target thickness(cm) | 8 | 12 | 18 | 26 |

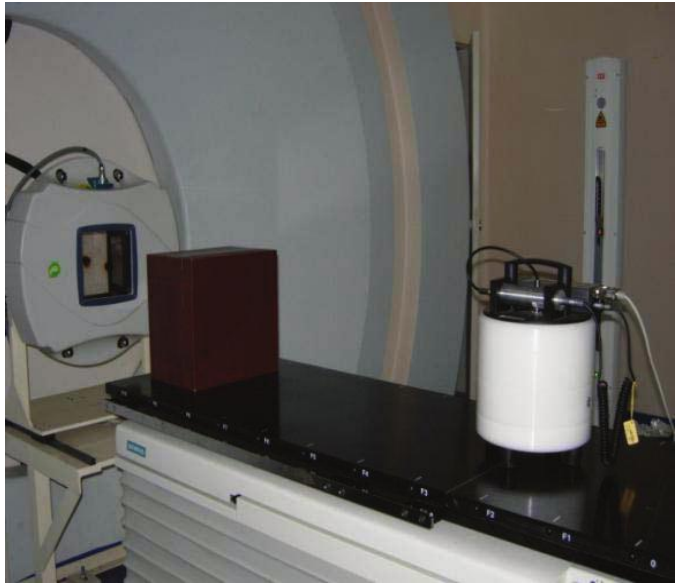
Fig. 2. Experimental arrangement.

**3 Monte Carlo simulation**

FLUKA code [14, 15] was used to simulate the neutron spectra and dose distributions. In our simulation, parallel $^{12}$C beam was delivered to the solid water target which has the same size and material as used in the experiment (table 1 and table 2). A set of detectors were placed at 1 meter far from the injection point at different angles as shown in Fig.3 in the Monte Carlo simulation of the neutron energy spectra. FLUKA version 2011.2b was used in this work.

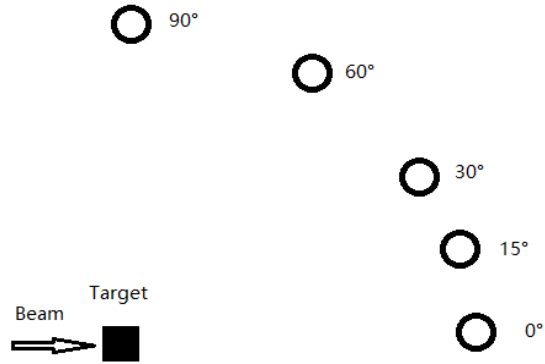

Fig. 3. The geometric arrangements for calculating the neutron energy spectra.

**4 Results and discussion**

Fig.4 shows the measured neutron dose as a function of the distance from the injection point to the detector for measurement angles of 0 degree, 30 degree and 90 degree, respectively. The data has been normalized with the beam intensity. It can be seen from Fig.4 that at ion energy of 350 MeV/u (corresponding to depth of 26 cm in tissue-like material), the maximum neutron dose is about 2.5 pSv/ion at 0 degree and the maximum data is about 0.015 pSv/ion at 90 degree. This is because in the small angle direction the cascade neutrons were the major part of the neutron yield, i.e., the neutron emission has a sharp forward peaking. In large angle, the measured neutrons are mainly from low energy evaporation mechanism. Fig.5 shows the comparison for the 0 degree experimental results with the FLUKA simulation. In this figure the relative unit was used, because the FLUKA simulation results were about three times larger than the experimental data. Fig.6 shows the neutron dose angular distributions, which were measured at 2 m far from the target. The relative unit was also used in this figure. In Fig.5 and Fig.6, the solid lines represent the FLUKA results, the points represent the measured results. It should be noted that there has about three times differences between the measurements and the FLUKA results. Several reasons could explain such differences qualitatively. First, the detector itself cannot achieve the ideal detector requirement, namely the detector energy response and human energy response to neutrons are not in absolute agreement. The second reason is related to the different geometric conditions between the simulation and the experiments. In experiments, the dose meter must have a slow body to stop the energetic neutrons, therefore the dose meter must have a definite volume. However, in simulation process a point detector is used throughout. Finally, the simulation process could not consider all the equipment in the therapy room. Therefor the contribution of the scattered neutrons from these equipment are not included.

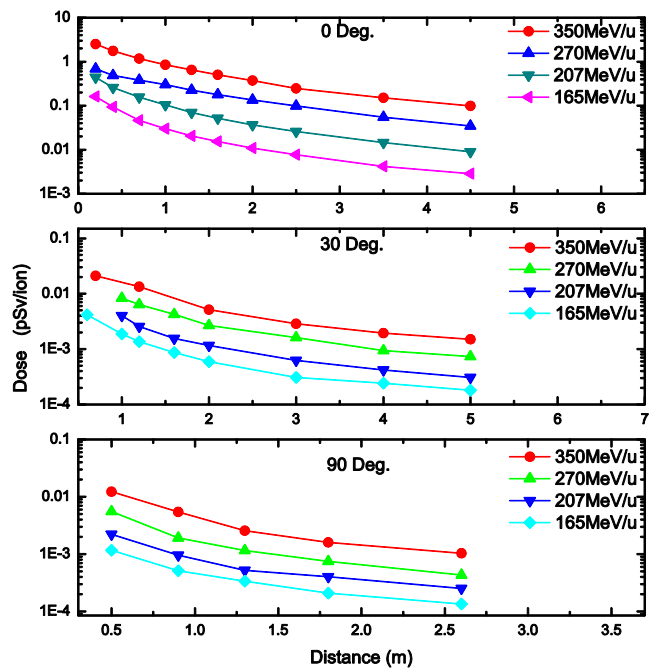

Fig. 4. Neutron dose distributions in the treatment room.

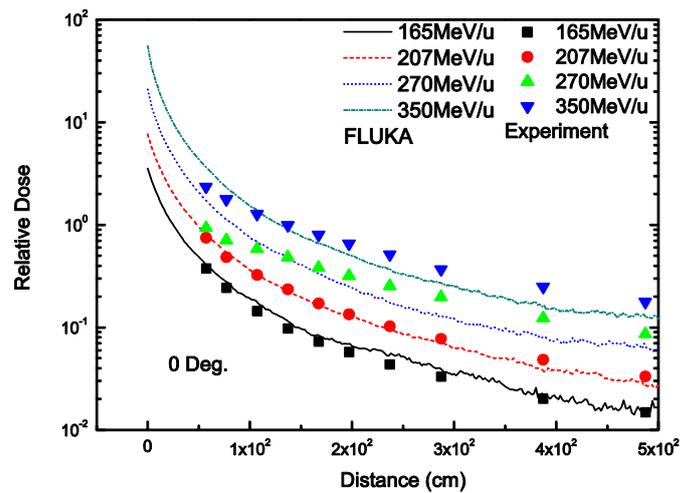

Fig. 5. The comparison for the 0 degree experimental results with the FLUKA simulation.

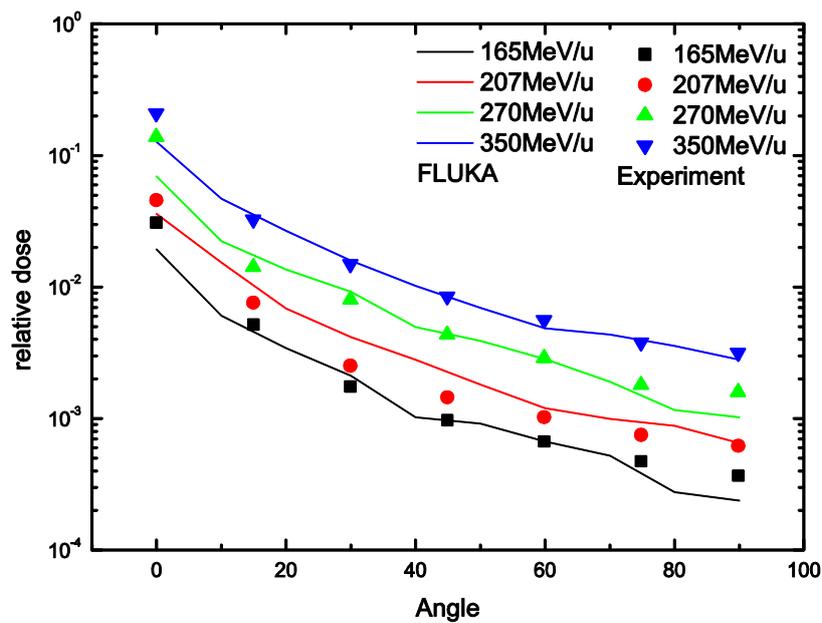

Fig. 6. The measured neutron dose angular distributions compared with the FLUKA simulation.

The neutron dose distribution cannot completely describe a neutron field. As described before, many factors may influence neutron dose in tissue. Thus studying neutron energy spectra is also necessary. Neutron dose can be calculated by using flux-dose conversion factor if neutron spectrum is known. Finally, the neutron energy spectra of $^{12}C$ ion bombarding on thick water target calculated by FLUKA code are shown in fig. 7, and Fig. 8 shows the corresponding neutron dose distribution.

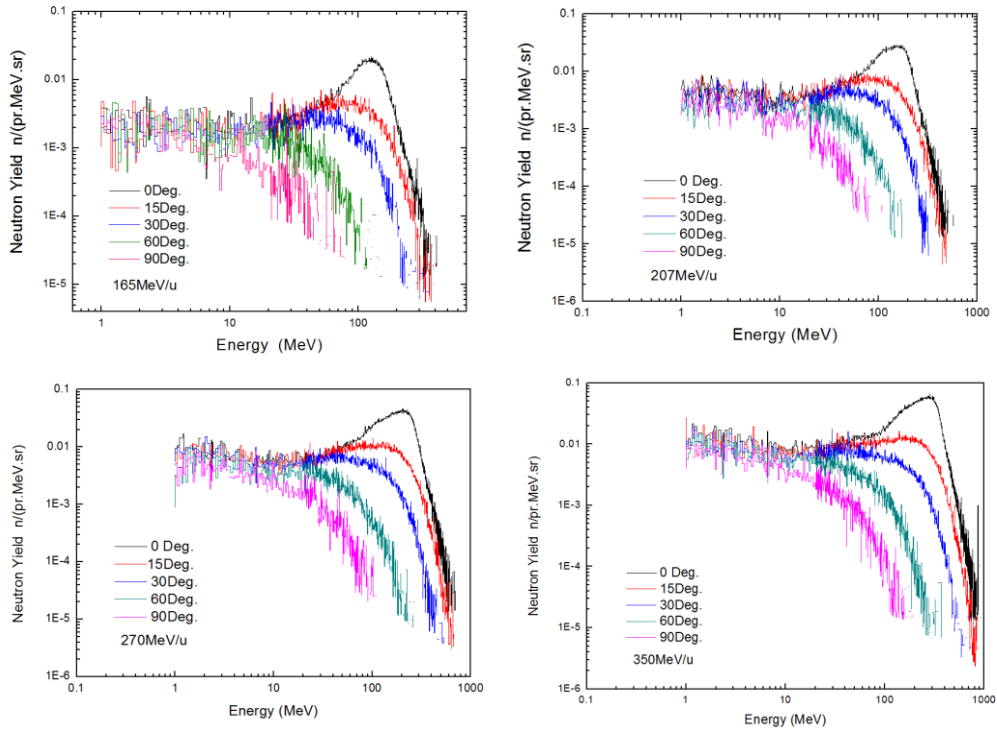

Fig. 7. The neutron energy spectra of $^{12}$C ion bombarding on thick water target.

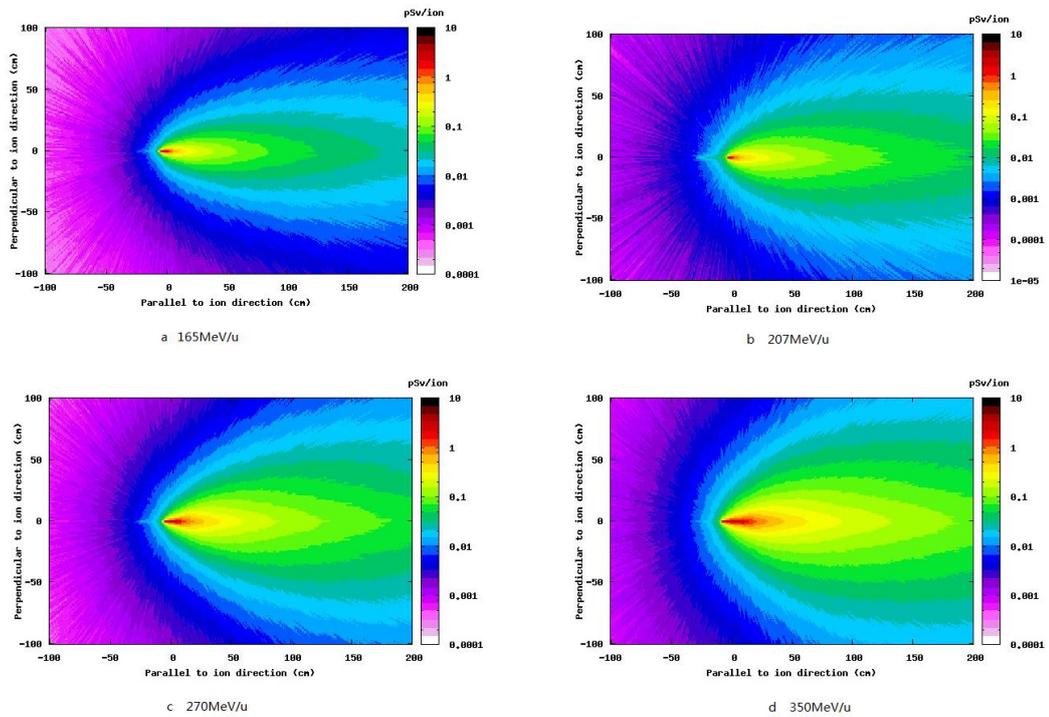

Fig. 8. Neutron dose distribution in the carbon ion therapy area.

## 5 Conclusions

In the present work, the neutron dose distribution on observation distances was measured at observation angle of 0 degree, 30 degree and 90 degree, respectively, as 165, 207, 270 and 350 MeV/u $^{12}$C ions bombarding on solid water target at the HIRFL deep tumor treatment terminal. We also simulated the neutron dose distribution by using the FLUKA code, and the results are in good agreement with experimental data. The secondary neutron energy spectra was also studied by using the FLUKA code. The experimental data is useful for shielding design of heavy ion accelerator facility and individual dose assessment.

## 6 Acknowledgment